\documentclass[twocolumn,showpacs,aps,prl,superscriptaddress]{revtex4}
\usepackage{graphicx,dcolumn,amsmath}
\usepackage{babarsym}

\newcommand{\BABARPubYear}    {05}
\newcommand{\BABARPubNumber}  {02}

\newcommand{\SLACPubNumber} {11037}

\def\scr{\scriptstyle}

\def\mMiss{\ensuremath{m_\mathrm{miss}}\xspace}

\def\dstar {\ensuremath{D^{\ast}}}
\def\dsstar{\ensuremath{D_s^{\ast}}}
\def\dstarm {\ensuremath{D^{\ast-}}}

\def\nBB {{\ensuremath{N_{\B\Bbar}}}}
\def\Dsp{\ensuremath{D_s^+}}

\def\bdstardsstar {\ensuremath{\Bz\to\Dstarm\Dss}\xspace}
\def\MC{{MC}\xspace}
\def\dsphipi{\ensuremath{\Ds\to\phi\pip}\xspace}
\let\Dsphipi=\dsphipi
\def\Kappa{\ensuremath{\mathcal{K}}}
\def\nDs{\ensuremath{\mathcal{N}_{D_{s}}}}
\def\nDsPhipi{\ensuremath{\mathcal{N}_{D_{s}\to\phi\pi}}}

\def\mmax{\ensuremath{m_\mathrm{max}}}

\def\makeatletter{\catcode `\@=11\relax}
\def\makeatother{ \catcode `\@=12\relax}
\makeatletter
\let\@@comma=\,       \def\,{\ensuremath{\@@comma}}
\let\@@colon=\;       \def\;{\ensuremath{\@@colon}}
\makeatother

\let\MeV=\mev
\let\GeV=\gev
\let\MeVc=\mevc
\let\GeVc=\gevc
\let\MeVcc=\mevcc
\let\GeVcc=\gevcc

\def\etal{\ensuremath{\textit{et al.}}\xspace}
\def\vev#1{\ensuremath{\langle #1\rangle}}

\long\def\inst#1{\par\nobreak\kern 4pt\nobreak
    {\it #1}\par\vskip 10pt plus 3pt minus 3pt}

\begin{document}

\begin{flushleft}
  SLAC-PUB-\SLACPubNumber \\
  \babar-PUB-\BABARPubYear/\BABARPubNumber
\end{flushleft}

\title{\large\bf
  Measurement of the \boldmath $B^0\to D^{\ast-} D_s^{\ast+}$ and
  $D_s^+\to\phi\pi^+$ Branching Fractions}
\date{\today}

%
\author{B.~Aubert}
\author{R.~Barate}
\author{D.~Boutigny}
\author{F.~Couderc}
\author{Y.~Karyotakis}
\author{J.~P.~Lees}
\author{V.~Poireau}
\author{V.~Tisserand}
\author{A.~Zghiche}
\affiliation{Laboratoire de Physique des Particules, F-74941 Annecy-le-Vieux, France }
\author{E.~Grauges-Pous}
\affiliation{IFAE, Universitat Autonoma de Barcelona, E-08193 Bellaterra, Barcelona, Spain }
% ...
\collaboration{The \babar\ Collaboration}
\noaffiliation

\hyphenation{else-where}

\begin{abstract}
We present measurements of the branching fractions $\BR(\bdstardsstar)$
and $\BR(\Dsphipi)$, based on $123\times10^6$ $\FourS\to\BB$ decays 
collected by the \babar\ detector at the PEP-II asymmetric-energy \epem \B
factory.
A partial reconstruction technique is used to measure $\BR(\bdstardsstar)$
and the decay chain is fully reconstructed to
measure the branching fraction product $\BR(\bdstardsstar)\times\BR(\Dsphipi)$.
Comparing these two measurements provides a model-independent
determination of the \Dsphipi\ branching fraction.
We obtain
$\BR(\bdstardsstar) = (1.88 \pm 0.09 \pm 0.17)\%$ and
$ \BR(\dsphipi)     = (4.81 \pm 0.52 \pm 0.38)\%$, where the first uncertainties
are statistical and the second systematic.
\end{abstract}

\maketitle

Published measurements of $\BR(\bdstardsstar)$\,\cite{pdg2004,BAD524} are
limited by the uncertainties on the \Ds partial decay rates. A substantial
improvement can therefore be obtained using a partial reconstruction
technique where the \Ds is not explicitly reconstructed. The measurement of
$\BR(\bdstardsstar)$ provides a test of the details of the
factorization assumption\,\cite{Beneke} in the relatively high $q^2$
regime\;\cite{luo}. Partial reconstruction in addition allows
an unbiased measurement of the \Dsphipi branching fraction, which has
important implications for a wide range of $D_s$ and \B physics, as most of
the $D_s$ decay branching fractions are normalized to it\;\cite{pdg2004}.
As an example, an improved measurement of $\BR(\Dsphipi)$ would reduce the
experimental uncertainty on the constraint on the Unitary Triangle
parameter $\gamma$ from the measurement of the \CP violating 
asymmetry in $\Bz\to D^{\ast\pm}\pi^\mp$ decays\;\cite{dstarpiPRL}.

We used $(123\pm1)\times10^6$ $\BB$ decays collected at the \pep2\
asymmetric-energy $\epem$ \B factory with the \babar\ detector, which is
described in detail elsewhere\;\cite{babarNIM}. We provide here a brief
description of the detector components relevant for this analysis.
Charged-particle trajectories are measured by a silicon vertex
tracker (SVT) and  a drift chamber (DCH) immersed in a 1.5\,T solenoidal
magnetic field. 
The five-layer SVT enables tracks with low transverse momentum to be
reconstructed. The energy and direction of photons and electrons are
measured by a CsI(Tl)-crystal electromagnetic calorimeter (EMC).
Charged-particle identification is obtained from the measurement of energy
loss in the tracking system, and from the measurement of the number and the
angle of Cherenkov photons in a ring-imaging Cherenkov detector (DIRC).

To study efficiencies and backgrounds and to validate the analysis we use
several event samples produced with a Monte Carlo (MC) simulation of the \babar\
detector based on GEANT4\;\cite{GEANT4} and reconstructed through the
same chain as the data.

The $\bdstardsstar\to(\Ds\gamma)(\Dzb\pim)$ decay\;\cite{ChargeConj}
is reconstructed using two different methods.
The first method combines the fully reconstructed \Dstarm\ decay with the photon
from the $\Dss\to\Ds\gamma$ decay, without explicit \Ds reconstruction.
Denoting the measured yield by \nDs, we can write:
\begin{equation}
\BR(\bdstardsstar) \equiv \BR_1=
\frac{\nDs}{\Kappa\sum_i (\eps_i \BR_i)}.
\end{equation}
Here
$\Kappa\equiv2\nBB f_{00}\BR(\Dss\to\Ds\gamma)\BR(\Dstarm\to\Dzb\pim)$,
\nBB\ is the number of \B-meson pairs,
$f_{00}=0.499\pm0.012$\;\cite{f00Paper} is the fraction of
$\FourS\to\BzBzb$ decays, $\BR_i$ is the branching
fraction of \Dzb decay mode $i$, $\eps_i$ is the efficiency for  partially
reconstructing the \Bz with a photon, a low momentum (``soft'') pion and a
\Dzb reconstructed in mode $i$.

The second method, based on full reconstruction of the \bdstardsstar\ decay
via \Dsphipi\ ($\phi\to\Kp\Km$), measures the branching fraction product
$\BR_2\equiv\BR(\bdstardsstar)\times\BR(\Dsphipi)$:
\begin{equation}
\BR_2 =
\frac{\nDsPhipi}{\Kappa\BR(\phi\to\Kp\Km)
\sum_i (\eps'_i \BR_i)},
\end{equation}
where \nDsPhipi\ is the number of reconstructed decays and
$\eps'_i$ is the efficiency for fully reconstructing the
\Bz, including reconstruction of $\phi\to\Kp\Km$. The \Dsphipi\ branching
fraction is measured from the $\BR_2/\BR_1$ ratio:
\begin{equation}
\label{eq:brdsphipi}
\BR(\dsphipi)=  \frac{\BR_2}{\BR_1} =
\frac{\nDsPhipi \sum_i (\eps_i \BR_i)}
{\nDs \BR(\phi\to\Kp\Km)  \sum_i (\eps'_i \BR_i)},
\end{equation}
where the factor \Kappa\ drops out. Although the efficiencies $\eps_i$ and
$\eps'_i$ are in general different, they include common factors and many
systematic uncertainties cancel in the ratio.

To extract the signal in partially reconstructed events,
we compute the ``missing mass'' recoiling against the $\Dstarm\gamma$
system, assuming that a $\Bz\to\Dstarm\gamma X$ decay took place:
\begin{equation}
\mMiss = \sqrt{(E_B - E_{\dstar} - E_{\gamma})^2 -
({\bf p}_B -\!{\bf p}_{\dstar} -\!{\bf p}_{\gamma})^2},
\label{eq:nostramm}
\end{equation}
where all quantities are defined in the \FourS center-of-mass (CM) frame.
While the photon and \Dstarm\ energies ($E_\gamma$, $E_{\dstar}$) and their
three-momenta (${\bf p}_{\gamma}$, ${\bf p}_{\dstar}$) are measured,
kinematical constraints are needed to determine the \B\ four-momentum
($E_B$, ${\bf p}_B$).
In order to do that we equate the \B-meson energy with $E_\mathrm{beam}$,
the beam energy in the CM frame, and
calculate the cosine of the opening angle $\vartheta_{B\dstar}$ between the
\B\ and the \dstarm\ momentum vectors 
 from 4-momentum conservation in the
\bdstardsstar\ decay. This leaves the azimuthal angle of the \B meson
around the \Dstarm\ direction as the only undetermined parameter in the
kinematics of the decay. \MC studies show that an arbitrary choice
of this angle (we fix $\cos\phi_{B\dstar} = 0$) introduces a negligible
spread (of the order of 1.5 \MeVcc) in the \mMiss distribution.
The \mMiss\ distribution of signal events peaks at the nominal \Dsp\
mass\;\cite{pdg2004} with a width of about 15\MeVcc.

\hyphenation{second}
We suppress unphysical $\Dstarm\gamma$ combinations by requiring
$|\cos\vartheta_{B\dstar}|\le1.2$ and events from 
$\epem\to$\uubar, \ddbar, \ssbar, \ccbar production
by requiring the ratio of the second to the zeroth Fox-Wolfram
moments\;\cite{R2} to be less than 0.3.

\Dstarm\ candidates are reconstructed in the $\Dzb\pim$ mode using 
\Dzb decays to $\Kp\pim$, $\Kp\pim\pipi$, $\Kp\pim\piz$ and
$\KS\pipi$, listed here in order of decreasing purity. The $\chi^2$
probabilities of both the \Dz\ and \Dstar\ vertex fits are required
to be greater than 1\%. The \Dstarm\ momentum in the \FourS frame   
must satisfy $1.4 < p_{\dstar} < 1.9\GeVc$.
We require the reconstructed mass of the \Dz\ to be within 3 
standard deviations ($\sigma_{m_{\Dz}}$) of the measured peak value, and
$Q_{\dstar} \equiv m_{\dstar}-m_{\Dz}-m_\pi$ to satisfy $Q_{\mathrm{lo}}
< Q_{\dstar} < Q_{\mathrm{hi}}$, where the choice of limits
$Q_{\mathrm{lo}}=4.10-5.20\MeVcc$ and $Q_{\mathrm{hi}}=6.80-
7.90\MeVcc$ around the nominal value $Q^\mathrm{\scr
PDG}_{\dstar}=5.851~\MeVcc$ depends on the \Dz decay mode.
Kaon identification is required in $\Kp\pim\piz$ and $\Kp\pim\pipi$ modes.
The \KS\ from the $\KS\pipi$ mode must have an invariant mass within 
15\MeVcc\ of the nominal \KS\ mass and a flight length greater than 3\,mm.

If more than one \Dstarm\ candidate is found, we first retain those that
have the \Dzb reconstructed in the decay mode with the highest expected
purity. If ambiguities persist at this stage, we choose the best
candidate based on the track quality of the soft pion and finally on the
minimum value of
$\chi^2 =
  [(Q_{\dstar} - Q^\mathrm{\scr PDG}_{\dstar})/\sigma_{Q_{\dstar}}]^2
+ [(m_{\Dz} - m^\mathrm{\scr PDG}_{\Dz})/\sigma_{m_{\Dz}}]^2$, where
$\sigma_{Q_{\dstar}}$ is the measured resolution on $Q_{\dstar}$.

Photon candidates are chosen from clusters of energy deposited in the
EMC that are not associated with any charged track.
The energy spectrum of photons from the $\Dss\to\Dsp\gamma$ decay
is rather soft ($E_\gamma\lsim0.4\GeV$) and this makes controlling the
background due to random photon associations one of the main challenges in
the analysis.
We require $E_\gamma>142\MeV$ and use the energy profile of the
cluster to refine the photon selection, requiring a minimum cluster lateral
moment\;\cite{LAT} of $0.016$, and a minimum Zernike moment
$A_{20}$\;\cite{zernike} of $0.82$. We also reject photon candidates that
form in combination with any other photon in the event a \piz\ whose
invariant mass is between 115 and 155\MeVcc\ and whose momentum in the CM
frame is greater than 200\MeVc.
This selection retains more than one photon candidate in about 10\,\% of
the events.
In these occurrences we choose the one that maximizes the value of a
likelihood ratio based on the energy and the shape of the reconstructed
cluster.

The cuts are chosen to maximize the expected statistical
significance of the selected signal using \MC. The combinatorial
background is dominated by \BzBzb events. 
None of the background components peak at the \Dsp\ mass in the \mMiss
distribution.
The reconstruction and selection efficiency, evaluated on simulated
events, is
$\vev{\eps\BR}\equiv\sum_i{(\eps_i \BR_i)}=(5.15 \pm
0.03)\times10^{-3}$.

\begin{figure}[!hbtp]
  \centering
  \includegraphics*[width=\hsize]{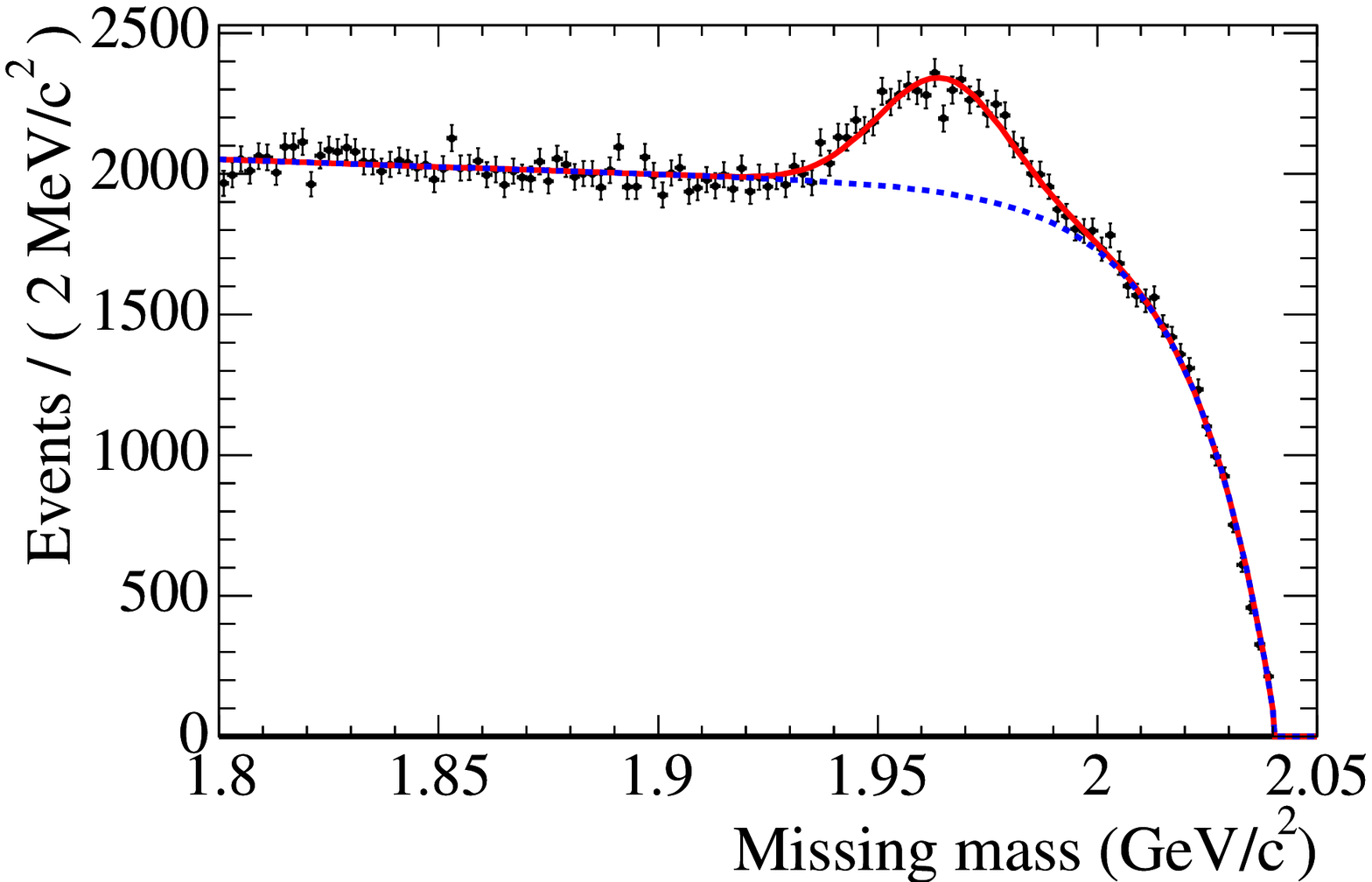}
  \caption{Fit (solid line) to the measured missing-mass distribu\-tion.
  The background component is shown as the dashed line.}
\label{fig:missmasstot}
\end{figure}

We extract the signal yield using an unbinned maximum-likelihood fit
to the \mMiss distribution. The signal peak is well
described by a Gaussian probability density function (p.d.f.). 
We parameterize the combinatorial background with the threshold function
$B(\mMiss)=B_0(1-e^{-(\mmax-\mMiss)/b})({\mMiss}/{\mmax})^c$.
Fig.~\ref{fig:missmasstot} shows the result of the fit to the missing-mass
distribution.
The width of the Gaussian signal distribution is taken
from \MC simulation.
The signal yield is $\nDs=7488 \pm 342$ events, corresponding to a
branching fraction $\BR(\bdstardsstar) = (1.88 \pm 0.09)\,\%$, where the
quoted error is purely statistical.

We now describe the full reconstruction of the $\bdstardsstar\to
(\Ds\gamma) (\Dzb\pim)$ chain, with $\Dzb$ decaying into the four modes
considered, and $\Ds\to\phi\pip\to K^+K^-\pip$.
Two kinematical variables are used: $\DeltaE \equiv
E_B-E_\mathrm{beam}$ and the energy-substituted mass
$\mes=\sqrt{E^2_{\mathrm{beam}}-{\bf p}^2_B}$.
The two variables have very little correlation; for signal events
\DeltaE peaks around zero and \mes at the \B-meson mass.
After applying selection cuts (described below) on the $\Dss$  and
$\Dstarm$ candidates, we retain the combination with the smallest value of
$|\DeltaE|$. The number of fully reconstructed \Bz\ candidates is then
obtained from a fit to the \mes spectrum.

The selection of $\dstarm$ candidates and most of the requirements on
photon candidates are identical to those adopted in the partial
reconstruction analysis.
Due to the additional kinematical constraints on fully reconstructed \B
decays, the combinatorial background level is much smaller; we can
therefore relax the requirement on
$E_\gamma$, thus improving the statistical significance of our sample.
We reconstruct $\phi$ candidates from pairs of oppositely charged tracks,
with at least one track satisfying kaon selection criteria;
\Ds\ candidates are formed by combination with an additional track, with
charge opposite to the soft pion from the \Dstarm\ decay. A mass within
$\pm$50\,\MeVcc\ of the nominal \Dsp\ mass\;\cite{pdg2004} is
required. Finally, \Dstarm\ and \Dss\ mass constraints are imposed in order
to improve the \mes and \DeltaE resolution of the \Bz candidate.
We require the $m_{\dsstar}-m_{D_s}$ mass difference to be between 125
and 160 \MeVcc, the reconstructed $\phi$ mass to be between 1.008 and
1.035\GeVcc, $E_\gamma$ to be greater than 90\MeV, and $|\Delta E|$ to
be less than 50\MeV.

\begin{figure}[htbp]
 \centering
  \includegraphics[width=0.95\hsize]{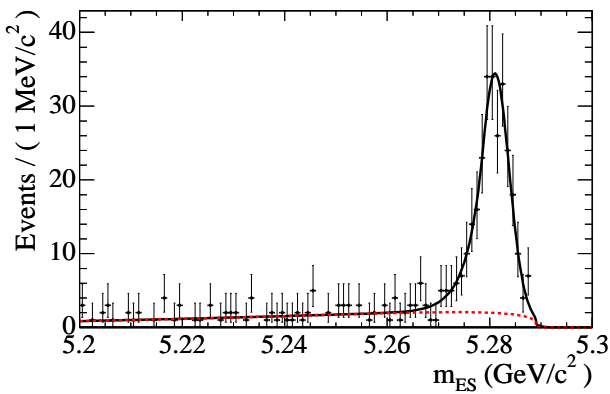}
  \caption{Fit (solid line) to the measured \mes distribution. The 
  background component is shown as the dashed line.}
\label{fig:fitdatafull}
\end{figure}
We perform an unbinned maximum-likelihood fit to the \mes\ distribution
with the sum of a Crystal Ball\;\cite{cbfunc} function, and a threshold
ARGUS\;\cite{Argus_bkgd} function; the latter accounts for the combinatorial
background.
From the fit to the data sample, shown in Fig.~\ref{fig:fitdatafull}, we
obtain $(247\pm19)$ events in the signal region defined as
$\mes>5.27\,\GeVcc$.

\MC studies indicate a peaking contribution due to real
$\bdstardsstar$ events, where either the \Dzb\ does not decay into the
reconstructed modes, or the \Ds\ does not decay into $\phi\pip$.
We subtract the peaking background applying a correction factor to take
into account that the values of the \bdstardsstar and \dsphipi
branching fractions that we have measured are different from those used in the
simulation, with an iterative procedure.
The resulting number of peaking background events expected in the data
sample is $35\pm6$ events; this uncertainty is taken into account in the
systematic error. After subtraction of the peaking background events, the
final signal event yield is $\nDsPhipi = (212 \pm 19)$.
Taking into account the reconstruction and selection efficiency
$\vev{\eps'\BR} \equiv\sum_i{(\eps'_i \BR_i)} = (6.16 \pm
0.24)\times10^{-3}$, evaluated on simulated events, we determine
$\BR_2=\BR(\bdstardsstar)\times\BR(\Dsphipi) = (8.81 \pm 0.86)
\times 10^{-4}$, where the error is statistical only.

\begin{table}[!htbp]
\begin{center}
\caption{Summary of systematic uncertainties.}
\label{tab:syst}
\begin{tabular}{||l | c | c | c ||}
\hline\hline
Source                           & ~$\BR_1$ [\%]~ & ~$\BR_2$ [\%]~  & ~$\BR_2/\BR_1$ [\%]~ \\ \hline
p.d.f.~modeling                  &    4.8  &           &  4.8  \\
Comb. background                 &         &  2.9      &  2.9  \\
\MC statistics                   &    0.6  &  3.2      &  3.3  \\
Peaking background               &         &  2.8      &  2.8  \\
\B\ counting                     &    1.1  &  1.1      &       \\
$f_{00}$                         &    2.4  &  2.4      &       \\
Soft pion efficiency             &    2.2  &  2.2      &       \\
\Dstarm\ Tracking efficiency     &    2.4  &  2.4      &       \\
\Dstarm\ Vertexing efficiency    &    2.0  &  2.0      &       \\
\Ds Tracking efficiency          &         &  2.6      &  2.6  \\
\Ds Vertexing efficiency         &         &  2.0      &  2.0  \\
Photon efficiency                &    1.8  &  1.8      &       \\
\piz\ eff. (\Dzb\to$\Kp\pim\piz$) &    1.2  &  1.2      &       \\
\piz\ veto                       &    4.7  &  4.7      &       \\
Particle identification          &    0.4  &  0.4      &       \\
Polarization uncertainty         &    0.8  &  0.8      &       \\
\Dz branch. fract.\;\cite{pdg2004}   &    3.2  &  3.2      &       \\
$\BR(\Dstarm\to\Dzb\pim)$\;\cite{pdg2004} &    0.7  &  0.7      &       \\
$\BR(\Dss\to\Ds\gamma)$\;\cite{hepex0408094} &    0.8  &  0.8      &       \\
$\BR(\phi\to\Kp\Km)$\;\cite{pdg2004}      &         &  1.2      &  1.2  \\
\hline
Total systematic error           &    9.1  & 10.7      & 7.9   \\
\hline\hline
\end{tabular}
\end{center}
\end{table}
The main sources of systematic uncertainties on the \bdstardsstar
branching fraction measurement are listed in the second column ($\BR_1$)
of Table~\ref{tab:syst}.
We compared the resolution of the Gaussian p.d.f.~in data and \MC
by fitting the missing mass distribution in the very clean sample of
fully reconstructed \bdstardsstar events. We disentangle in this way
the effect of the experimental resolution on the width of the signal peak
from the correlations in the fit between the width and the background
parameters. We obtain
$\sigma_\mathrm{data}/\sigma_\mathrm{MC}~=~(1.01\pm 0.05)$.
We repeated the \mMiss fits changing the Gaussian width by this
uncertainty, and varying the background parameters by their errors. We
assign the maximum deviation as systematic uncertainty, labelled in
Table~\ref{tab:syst} as ``p.d.f.~modeling''.
The \MC statistics uncertainty is the statistical error
on the efficiency determination.
The systematic uncertainties due to tracking, vertexing, photon and \piz
reconstruction efficiencies, and particle identification are evaluated using
independent control samples. The effect of the \piz\ veto is
evaluated from fully reconstructed events.
The uncertainty due to the dependence of the efficiency on the polarization
of the \bdstardsstar decay is assessed from \MC samples generated with complete
longitudinal and transverse polarization.
In the full reconstruction analysis the error on peaking background is due
to the \MC statistics and to the uncertainty on the relevant \Dzb\ and \Ds\
branching fractions; the uncertainty on the combinatorial background is
estimated using the \DeltaE\ sideband ($|\DeltaE|>200\MeV$) as an
alternative way of computing the number of background events under the \mes
peak. Several systematic uncertainties in the full reconstruction are in
common with the partial reconstruction analysis, and therefore cancel in
the ratio of Eq.~\ref{eq:brdsphipi}.
All remaining sources are listed in the last column of
Table\;\ref{tab:syst}.

We repeated both the partial and the full reconstruction analyses on
generic \MC samples consisting of \BzBzb, \BpBm, and low-mass
\qqbar events, finding no bias. The result is also stable over different
data-taking periods.

In summary, we have measured the \bdstardsstar\ branching fraction 
\begin{equation}
 \BR(\bdstardsstar) = (1.88 \pm 0.09 \pm 0.17)\%,
\end{equation}
where the first uncertainty is statistical and the second is
systematic. This result is independent of the partial
decay rates of the \Ds\ mesons.
It is consistent with a previous \babar\ measurement\;\cite{BAD524} and
with the world average, and reduces the total uncertainty by a factor of
about three.
The measurement is in agreement with the predictions of the factorization
model $\BR(\bdstardsstar)_\mathrm{theor}=(2.4\pm0.7)\,\%\;$\cite{luo}.

We have measured the branching fraction of \Dsphipi\ decay:
\begin{equation}
  \BR(\dsphipi) = (4.81 \pm 0.52 \pm 0.38)\%.
\end{equation}
This result represents an improvement by about a factor of two over previous
measurements\;\cite{CLEO,pdg2004}.

We are grateful for the excellent luminosity and machine conditions
provided by our \pep2\ colleagues, 
and for the substantial dedicated effort from
the computing organizations that support \babar.
The collaborating institutions wish to thank 
SLAC for its support and kind hospitality. 
This work is supported by
DOE
and NSF (USA),
NSERC (Canada),
IHEP (China),
CEA and
CNRS-IN2P3
(France),
BMBF and DFG
(Germany),
INFN (Italy),
FOM (The Netherlands),
NFR (Norway),
MIST (Russia), and
PPARC (United Kingdom). 
Individuals have received support from CONACyT (Mexico), A.~P.~Sloan Foundation, 
Research Corporation,
and Alexander von Humboldt Foundation.


\begin{thebibliography}{99}
\bibitem{ChargeConj}
Charge conjugate decays are included implicitly throughout the paper.
\end{thebibliography}
\end{document}